\documentclass[aip,amsmath,amssymb,twocolumn,]{revtex4}                

\usepackage{textcase}
\usepackage{graphicx}
\usepackage{dcolumn}
\usepackage{bm}
\usepackage{amsmath,amsthm,amssymb,mathrsfs}

\begin{document}


\title{Current dependence of spin torque switching rate based on Fokker-Planck approach}

\author{Tomohiro Taniguchi}
\author{Hiroshi Imamura}

\affiliation{ 
National Institute of Advanced Industrial Science and Technology (AIST), Spintronics Research Center, Tsukuba 305-8568, Japan
}

\date{\today}%


\begin{abstract}
  The spin torque switching rate of an in-plane magnetized system 
  in the presence of an applied field 
  is derived by solving the Fokker-Planck equation. 
  It is found that 
  three scaling currents are necessary to describe 
  the current dependence of the switching rate 
  in the low-current limit. 
  The dependences of these scaling currents on the applied field strength are also studied. 
\end{abstract}

\maketitle


Spin torque induced magnetization switching of a nanostructured ferromagnet 
in the thermally activated region is an important phenomenon 
for spintronics applications 
because the thermal stability and the spin torque switching current 
of the magnetic random access memory (MRAM) 
can be obtained from its switching probability \cite{albert02,myers02,morota08,yakata09,bedau10}. 
The experimentally observed switching probability has been analyzed by the formula 
$P=1-e^{-\nu t}$ \cite{brown63,koch04,li04,apalkov05,suzuki09,butler12,pinna12,pinna13,kalmykov13,taniguchi11a,taniguchi12,taniguchi12b,taniguchi13,taniguchi13b}, 
where the switching rate $\nu=f e^{-\Delta}$ consists of 
attempt frequency $f$ and switching barrier $\Delta$. 
It has been often assumed that 
the attempt frequency $f$ is constant (typically 1 GHz \cite{yakata09}), 
and that the switching barrier is proportional to current as $\Delta=\Delta_{0}(1-I/I_{\rm c}^{*})$, 
where the thermal stability $\Delta_{0}=MH_{\rm K}V/(2k_{\rm B}T)$ 
consists of magnetization $M$, 
uniaxial anisotropy field along the easy axis $H_{\rm K}$, 
volume of the free layer $V$, 
and temperature $T$. 
The current is denoted as $I$ while $I_{\rm c}^{*}$ is the spin torque switching current 
at zero temperature. 


However, our recent works revealed the limitation 
of the applicability of the previous theories \cite{taniguchi11a,taniguchi12b,taniguchi13,taniguchi13a}. 
For example, the value of the attempt frequency depends on the current magnitude. 
Also, the linear scaling of the switching barrier $\Delta$ 
is valid only for $I<I_{\rm c}$, 
while $\Delta$ depends on the current nonlinearly for $I_{\rm c} \le I < I_{\rm c}^{*}$, 
where $I_{\rm c}(<I_{\rm c}^{*})$ is a characteristic current of the instability of the equilibrium state. 
The formula in Refs. \cite{taniguchi13,taniguchi13a} will enable us 
to evaluate the thermal stability and the switching current 
with high accuracies. 
However, Refs. \cite{taniguchi13,taniguchi13a} consider only the zero applied field case, 
while in the experiments the applied field has been often used to quickly observe the switching \cite{albert02,myers02,yakata09,bedau10}. 


In this paper, 
we derive the theoretical formula of the switching rate 
of an in-plane magnetized system 
in the presence of the applied field 
by applying the mean first passage time approach to 
the Fokker-Planck equation. 
We find that in the low-current region ($I<I_{\rm c}$), 
the current dependence of the switching rate is 
characterized by three scaling currents, $I_{\rm c}$, $\tilde{I}_{\rm c}$, and $I_{\rm c}^{*}$. 
The applied field dependences of these scaling currents are also studied. 


\begin{figure}
\centerline{\includegraphics[width=0.7\columnwidth]{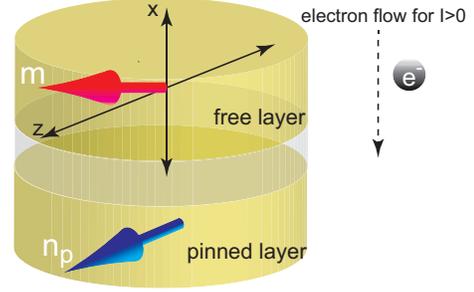}}\vspace{-3.0ex}
\caption{
         Schematic view of an in-plane magnetized system. 
         \vspace{-3ex}}
\label{fig:fig1}
\end{figure}



The system we consider is schematically shown in Fig. \ref{fig:fig1}, 
where the unit vectors pointing in the magnetization directions of the free and the pinned layers are 
denoted as $\mathbf{m}$ and $\mathbf{n}_{\rm p}=\mathbf{e}_{z}$, respectively. 
The $z$-axis is parallel to the in-plane easy axis of the free layer 
while the $x$-axis is normal to the film-plane. 
The positive current is defined as the electron flow from the free layer to the pinned layer. 
The energy density of the free layer, 
\begin{equation}
  E
  =
  -M H_{\rm appl}
  m_{z}
  -
  \frac{MH_{\rm K}}{2}
  m_{z}^{2}
  +
  \frac{4\pi M^{2}}{2}
  m_{x}^{2},
  \label{eq:energy}
\end{equation}
consists of 
the Zeeman energy, 
the uniaxial anisotropy energy along the $z$-axis, 
and the shape anisotropy along the $x$-axis, respectively. 
The minima of the energy density are 
$\mp M H_{\rm appl}-(MH_{\rm K}/2)$, 
corresponding to $\mathbf{m}=\pm \mathbf{e}_{z}$, 
while the energy density at the saddle point, $\mathbf{m}=(0,\pm\sqrt{1-(H_{\rm appl}/H_{\rm K})^{2}},-H_{\rm appl}/H_{\rm K})$, is 
$E_{\rm s}=MH_{\rm appl}^{2}/(2H_{\rm K})$. 
Below, the initial state is taken to be $\mathbf{m}=\mathbf{e}_{z}$. 
The applied field magnitude $|H_{\rm appl}|$ should be less than $H_{\rm K}$ 
to guarantee two minima of $E$. 
The magnetization dynamics is described by 
the Landau-Lifshitz-Gilbert (LLG) equation, 
\begin{equation}
  \frac{d \mathbf{m}}{dt}
  =
  -\gamma
  \mathbf{m}
  \times
  \mathbf{H}
  -
  \gamma
  H_{\rm s}
  \mathbf{m}
  \times
  \left(
    \mathbf{n}_{\rm p}
    \times
    \mathbf{m}
  \right)
  +
  \alpha
  \mathbf{m}
  \times
  \frac{d \mathbf{m}}{dt},
  \label{eq:LLG}
\end{equation}
where $\mathbf{H}=-\partial E/\partial (M \mathbf{m})$. 
The gyromagnetic ratio and the Gilbert damping constant are 
denoted as $\gamma$ and $\alpha$, respectively. 
The spin torque strength, 
\begin{equation}
  H_{\rm s}
  =
  \frac{\hbar \eta I}{2eMV},
  \label{eq:H_s}
\end{equation}
includes the spin polarization $\eta$ of the current. 


At zero temperature, 
the initial state becomes unstable 
when the current magnitude is larger than 
\begin{equation}
  I_{\rm c}
  =
  \frac{2\alpha eMV}{\hbar \eta}
  \left(
    H_{\rm appl}
    +
    H_{\rm K}
    +
    2\pi M
  \right).
  \label{eq:Ic}
\end{equation}
The instability of the initial state does not guarantee the switching. 
The switching at zero temperature 
occurs when the current magnitude becomes larger than \cite{hillebrands06} 
\begin{equation}
  I_{\rm c}^{*}
  =
  \frac{2\alpha eMV}{\hbar\eta}
  4\pi M 
  \frac{\mathscr{N}}{\mathscr{D}}.
  \label{eq:Ic*}
\end{equation}
Here, $\mathscr{N}$ and $\mathscr{D}$ are defined as 
\begin{equation}
\begin{split}
  &
  \mathscr{N}
  =
\\
  &
  \sqrt{
    1 \!+\! k
  }
  [k(1 \!+\! k) \!-\! h^{2}]
  \left\{
    2(k^{2} \!-\! h^{2})
    \sqrt{
      k(1 \!+\! k)
    }
  \right.
\\
  & +
  \left.
    h 
    \sqrt{
      k(k^{2} \!-\! h^{2})
    }
    \left[
      \pi
      +
      2 \sin^{-1}
      \left(
        \!
        \frac{h}{\sqrt{k[k(1+k) \!-\! h^{2}]}}
        \!
      \right)
    \right]
  \right\},
\end{split}
\end{equation}
\begin{equation}
\begin{split}
  &
  \mathscr{D}
  =
\\
  &
  2 h \sqrt{k}
  (1 \!+\! k)
  (k^{2} \!-\! h^{2})
  \!+\!
  k[k(1 \!+\! k) \!-\! h^{2}]
  \sqrt{
    k(1 \!+\! k)
    (k^{2} \!-\! h^{2})
  }
\\
  & 
  \times
  \!\!
  \left[
    \pi
    +
    2 \sin^{-1}
    \left(
      \!
      \frac{h}{\sqrt{k[k(1+k) \!-\! h^{2}]}}
      \!
    \right)
  \right],
\end{split}
\end{equation}
where $h=H_{\rm appl}/(4\pi M)$ and $k=H_{\rm K}/(4\pi M)$. 


In the thermally activated region $I<I_{\rm c}^{*}$, 
the magnetization dynamics is described by the Fokker-Planck equation, 
which can be obtained 
by adding the stochastic torque, $-\gamma \mathbf{m}\times\mathbf{h}$, to 
the right hand side of Eq. (\ref{eq:LLG}) \cite{brown63}, 
and is given by \cite{apalkov05,dykman12} 
\begin{equation}
  \frac{\partial \mathcal{P}}{\partial t}
  +
  \frac{\partial J}{\partial E}
  =
  0,
  \label{eq:Fokker-Planck}
\end{equation}
\begin{equation}
\begin{split}
  J(E)
  =&
  \frac{M}{\gamma}
  \left(
    \frac{\mathscr{M}_{\rm s}-\alpha \mathscr{M}_{\alpha}}{\tau}
    +
    \frac{MD \mathscr{M}_{\alpha}}{\gamma \tau^{2}}
    \frac{d \tau}{dE}
  \right)
  \mathcal{P}
\\
  &-
  D 
  \left(
    \frac{M}{\gamma}
  \right)^{2}
  \frac{\mathscr{M}_{\alpha}}{\tau}
  \frac{\partial \mathcal{P}}{\partial E},
  \label{eq:current}
\end{split}
\end{equation}
where $\mathcal{P}$ and $J$ are 
the probability function of the magnetization direction 
and the probability current, respectively. 
The diffusion constant $D=\alpha\gamma k_{\rm B}T/(MV)$ relates 
to the fluctuation-dissipation theorem as 
$\langle h_{i}(t) h_{j}(t^{\prime}) \rangle = (2D/\gamma^{2}) \delta_{ij} \delta(t-t^{\prime})$. 
The functions 
$\mathscr{M}_{\rm s}(E)=\gamma^{2}H_{\rm s} \oint dt [\mathbf{n}_{\rm p}\cdot\mathbf{H}-(\mathbf{m}\cdot\mathbf{n}_{\rm p})(\mathbf{m}\cdot\mathbf{H})]$ 
and $\mathscr{M}_{\alpha}(E)=\gamma^{2} \oint dt [\mathbf{H}^{2} - (\mathbf{m}\cdot\mathbf{H})^{2}]$ 
are proportional to the work done by spin torque 
and the energy dissipation due to the damping 
on constant energy line, respectively. 
The precession period on the constant energy line is denoted as $\tau$. 
Equation (\ref{eq:Fokker-Planck}) 
describes the Brownian motion of the magnetization 
in the effective potential $\mathscr{E}$ defined as 
\begin{equation}
  \mathscr{E}
  =
  \int^{E} 
  d E^{\prime}
  \left[
    1
    -
    \frac{\mathscr{M}_{\rm s}(E^{\prime})}{\alpha \mathscr{M}_{\alpha}(E^{\prime})}
  \right].
  \label{eq:effective_energy}
\end{equation}
The steady state solution of Eq. (\ref{eq:Fokker-Planck}) is proportional to $e^{-\mathscr{E}V/(k_{\rm B}T)}$. 
It should be noted that 
$I_{\rm c}$ and $I_{\rm c}^{*}$ satisfy 
$\lim_{E \to -MH_{\rm appl}-(MH_{\rm K}/2)}d \mathscr{E}/d E=1-I/I_{\rm c}$ 
and 
$\lim_{E \to E_{\rm s}}d \mathscr{E}/dE =1-I/I_{\rm c}^{*}$, respectively. 



\begin{figure}
\centerline{\includegraphics[width=0.8\columnwidth]{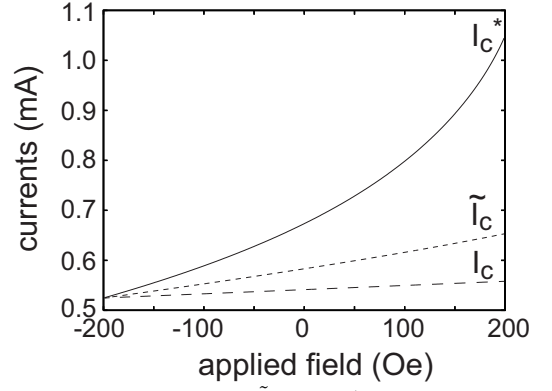}}\vspace{-3.0ex}
\caption{
         Dependences of $I_{\rm c}$, $\tilde{I}_{\rm c}$, and $I_{\rm c}^{*}$ 
         on the applied field. 
         \vspace{-3ex}}
\label{fig:fig2}
\end{figure}



The mean first passage time \cite{hanggi90}, 
which characterizes how long the magnetization stays 
in the stable region of the effective potential $\mathscr{E}$, 
can be introduced as $\mathcal{T}= \int_{0}^{\infty} dt \int_{E^{*}}^{E_{\rm s}} d E_{1} \mathcal{P}(E_{1},t|E^{*},0)$. 
Here, $E^{*}$ for $I<I_{\rm c}$ is the energy density at the initial state, $-MH_{\rm appl}-(MH_{\rm K}/2)$, 
while $E^{*}$ for $I_{\rm c} \le I < I_{\rm c}^{*}$ is determined by 
the condition $\mathscr{M}_{\rm s}(E^{*})=\alpha \mathscr{M}_{\alpha}(E^{*})$. 
The solution of the mean first passage time is obtained from Eq. (\ref{eq:Fokker-Planck}), 
and is given by 
\begin{equation}
\begin{split}
  \mathcal{T}
  =
  \frac{\gamma V}{\alpha M k_{\rm B}T}
  &
  \int_{E^{*}}^{E_{\rm s}} 
  d E_{1}
  \int_{E^{*}}^{E_{1}} 
  d E_{2}
  \frac{\tau(E_{2})}{\mathscr{M}_{\alpha}(E_{1})}
\\
  &
  \times
  \exp
  \left\{
    \frac{[ \mathscr{E}(E_{1})-\mathscr{E}(E_{2})]V}{k_{\rm B}T}
  \right\}.
  \label{eq:MFPT}
\end{split}
\end{equation}
The switching rate from $\mathbf{m}=\mathbf{e}_{z}$ to $\mathbf{m}=-\mathbf{e}_{z}$ is 
given by $\nu=(1+I/I_{\rm c}^{*})/(2 \mathcal{T})$ \cite{taniguchi13a}. 
In the low-current region $I < I_{\rm c}$ 
and in the high-barrier limit, 
the switching rate is 
\begin{equation}
\begin{split}
  \nu
  =&
  \frac{\alpha MV \mathscr{M}_{\alpha}(E_{\rm s})}{2 \gamma k_{\rm B}T \tau(E^{*})}
  \left(
    1
    -
    \frac{I}{I_{\rm c}}
  \right)
  \left[
    1
    -
    \left(
      \frac{I}{I_{\rm c}^{*}}
    \right)^{2}
  \right]
\\
  & 
  \times
  \exp
  \left[
    -\Delta_{0}
    \left(
      1
      +
      \frac{H_{\rm appl}}{H_{\rm K}}
    \right)^{2}
    \left(
      1
      -
      \frac{I}{\tilde{I}_{\rm c}}
    \right)
  \right],
  \label{eq:rate}
\end{split}
\end{equation}
where $\mathscr{M}_{\alpha}(E_{\rm s})=8 \pi \gamma M \mathscr{N}/[k^{2}(1+k)^{2}\sqrt{k^{2}-h^{2}}]$ 
and $\tau(E^{*})=2\pi/[\gamma \sqrt{(H_{\rm appl}+H_{\rm K})(H_{\rm appl}+H_{\rm K}+4\pi M)}]$. 
The scaling current $\tilde{I}_{\rm c}$ is defined as 
\begin{equation}
  \tilde{I}_{\rm c}
  =
  \frac{2\alpha eMV}{\hbar\eta}
  \frac{4\pi M}{\mathcal{S}}, 
  \label{eq:Ic_tilde}
\end{equation}
where the dimensionless quantity $\mathcal{S}$ is defined as 
\begin{equation}
  \mathcal{S}
  =
  \frac{4\pi M \int_{E^{*}}^{E_{\rm s}} d E^{\prime} \mathscr{M}_{\rm s}(E^{\prime})/\mathscr{M}_{\alpha}(E^{\prime})}{(MH_{\rm K}/2)(1+H_{\rm appl}/H_{\rm K})^{2}H_{\rm s}}. 
\end{equation}
As mentioned above, 
Eq. (\ref{eq:rate}) is valid 
for $I<I_{\rm c}$ 
and $\Delta=\Delta_{0}(1+H_{\rm appl}/H_{\rm K})^{2}(1-I/\tilde{I}_{\rm c}) \gg 1$. 
The attempt frequency is defined as 
$f=\nu e^{\Delta}$. 
On the other hand, 
in the high-current region $I_{\rm c} \le I < I_{\rm c}^{*}$, 
the numerical calculation is necessary 
to estimate the current dependence of the switching rate $\nu$ \cite{taniguchi13a}. 


Figure \ref{fig:fig2} shows 
the dependences of 
$I_{\rm c}$, $\tilde{I}_{\rm c}$, and $I_{\rm c}^{*}$ 
on the applied field $H_{\rm appl}$. 
The values of the parameters are 
$M=1000$ emu/c.c., 
$H_{\rm K}=200$ Oe, 
$V = \pi \times 80 \times 35 \times 2.5$ nm${}^{3}$, 
$\alpha=0.01$, 
and $\eta=0.8$, respectively, 
which are typical values for a magnetic tunnel junction 
consisting of CoFeB \cite{morota08,yakata09,kubota05a,kubota05b}. 
The scaling current $\tilde{I}_{\rm c}$ is less than $I_{\rm c}^{*}$, 
and weakly depends on $H_{\rm appl}$, compared with $I_{\rm c}^{*}$. 



\begin{figure}
\centerline{\includegraphics[width=1.0\columnwidth]{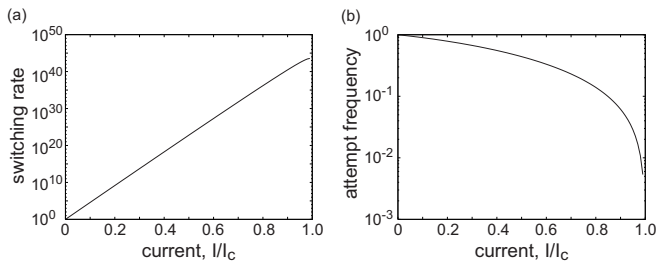}}\vspace{-3.0ex}
\caption{
         Dependences of (a) the switching rate $\nu$ and (b) the attempt frequency $f$ 
         in the low-current region ($I<I_{\rm c}$) 
         on the current $I$ for $H_{\rm appl}=100$ Oe, 
         where the values of $\nu$ and $f$ are normalized by 
         those at $I=0$ 
         while $I$ is normalized by $I_{\rm c}$
         \vspace{-3ex}}
\label{fig:fig3}
\end{figure}




Figures \ref{fig:fig3} (a) and (b) show 
the current dependence of the switching rate $\nu$ 
and the attempt frequency $f$ 
in the low-current region, $I<I_{\rm c}$. 
The applied field strength is $H_{\rm appl}=100$ Oe. 
The current dependence of $\nu$ in the logarithmic scale is approximately linear 
due to the linear dependence of the switching barrier $\Delta$, 
although the attempt frequency $f$ also depends on the current. 
It should be noted that the scaling current of the switching barrier 
in the low-current region is $\tilde{I}_{\rm c}$, 
neither $I_{\rm c}$ nor $I_{\rm c}^{*}$ as argued in Refs. \cite{koch04,li04}. 
This means that 
the previous analyses of the experiments \cite{morota08,yakata09} 
underestimate the switching current. 


In the above formula, 
the effect of the field like torque \cite{suzuki09} is neglected. 
On the other hand, recently, 
the effect of the field like torque on the relaxation time, $1/\nu$, was 
experimentally investigated \cite{rippard11}, 
in which the field like torque term is treated 
as an additional applied field, 
and is assumed a quadratic function of the bias voltage. 
From the bias voltage dependence of the relaxation time, 
the expansion coefficient of the field like torque term was estimated. 
However, in Ref. \cite{rippard11}, 
the attempt frequency ($1/\tau_{0}$ in Ref. \cite{rippard11}) is assumed to be 
independent of the damping constant, temperature, and bias voltage. 
The combination of our formula developed above 
with the method in Ref. \cite{rippard11} will help 
the quantitative estimation of the retention time of MRAM with high accuracy. 


In summary, 
the theoretical formula of 
the spin torque switching rate of an in-plane magnetized system 
in the presence of an applied field was derived 
by solving the Fokker-Planck equation. 
In the low-current region $I<I_{\rm c}$, 
the current dependence of the switching rate is characterized by 
three scaling currents, $I_{\rm c}$, $\tilde{I}_{\rm c}$, and $I_{\rm c}^{*}$, 
where $I_{\rm c}$ and $I_{\rm c}^{*}$ determines the current dependence of the attempt frequency 
while $\tilde{I}_{\rm c}$ determines that of the switching barrier. 
The dependences of these scaling currents 
on the applied field strength were also studied. 


The authors would like to acknowledge 
H. Kubota, H. Maehara, K. Yakushiji, A. Fukushima, K. Ando, and S. Yuasa 
for the valuable discussions they had with us. 
This work was supported by JSPS KAKENHI Grant-in-Aid for Young Scientists (B) 25790044. 




\end{document}